\documentstyle[12pt,aps,epsf]{revtex}

\input{epsf}
\begin{document}

\title{Delocalised states in 1D diagonally disordered system}
\author{Kozlov G.G.}

\maketitle
\vskip20pt
\hskip100pt {\it e}-mail:  gkozlov@photonics.phys.spbu.ru
\vskip20pt
\begin{abstract}
1D diagonally disordered chain with Frenkel exciton and long range exponential
intersite interaction is considered. It is shown that some states of this
disordered system are delocalised contrary to the popular statement that all
states in 1D disordered system are localised.
\end{abstract}

\section{Introduction and main result}

It is well known that all wave functions of translationary symmetric
systems are delocalized.
One of the most interesting properties of the homogeneous disordered
  systems is the
possibility of localised wave functions.

The mathematical problems of the theory of  disordered systems are very
 complicated and for this reason the theory of disordered systems is not
 so well developed as the theory of symmetric systems.
Despite this fact some statements related to disordered systems are
 considered to be well established and reliable.
   The above mentioned   occurrence of the localised band is
 one of them. The next example of statement of this kind is that  all
 states in 1D disordered system are localised \cite{Pastur}.
Recently appeared the reports \cite{VM,AM} about the delocalisation in
1D systems with intersite interaction in the form:
$J_{n,m}=J/|n-m|^\nu, 1<\nu<3/2$. In this letter we consider 1D
   diagonally disordered chain
with exponential intersite interaction and present arguments (computer
simulations and theoretical treatment) in favor of partial delocalisation in this system.
In this section we describe the system and present numerical results
and in the next section we review the reasons which made us to study
this system and present the approximate expression for the mobility edge.

Let us consider 1D Frenkel exciton in diagonally disordered chain. The
mathematical problem is redused to the following random matrix of the
Hamiltonian:
\begin{equation}
 H_{r,r'}=\varepsilon_r\hskip2pt\delta_{r,r'}+w(r-r') \hskip10mm r,r'=1,...,N,
 \end{equation}

where

\begin{equation}
 w(r)=v_0\exp-|r/R|,
 \end{equation}

Random values $\varepsilon_r$ are supposed
 to be independent  and having the
distribution function:
\begin{equation}
\rho(\varepsilon)=\cases{1/\Delta \hskip3mm \varepsilon\in[0,\Delta]\cr
0 \hskip3mm \hbox{other cases}}
\hskip5mm\hbox{(Anderson's model)}
\end{equation}

The thermodynamic limit $N\rightarrow\infty$ is implied.
 To separate the localised and delocalised wave functions one should
 use some criterion of localisation.
 We use {\it the number of sites covered by the wave function } \cite{Kozlov}
 determined as follows.
 Let us consider some eigen vector $\Psi$ of the
 Hamiltonian (1) with components $\Psi_r, r=1,...,N$.
 What contribution one should ascribe to the arbitrary site $r$? It
 is naturally to accept that this contribution is zero if $|\Psi_r|^2=0$
 and equal to unit if $|\Psi_r|^2=$max$\{|\Psi_1|^2,...,|\Psi_N|^2\}$.
 So we come to the conclusion that the contribution of the arbitrary
 site $r$ is $|\Psi_r|^2/|\Psi|_{max}^2$.  The total number
  of sites $n(\Psi)$ covered by normalised eigen function $\Psi$ is
   the sum of contributions of all sites:

\begin{equation}
  n(\Psi)=\sum_{r=1}^N {|\Psi_r|^2\over|\Psi|^2_{max}}= {1\over|\Psi|^2_{max}}
  \end{equation}

Delocalisation in  (1),(2),(3)
appear when $R>>1$. Below we study the properties of the eigen
vectors of (1)  with $R=20, v_0=0.5, \Delta=4, N=1000$.

 The dependance of  number of sites covered by the wave
function against  corresponding energy for the Hamiltonian (1)
 is presented on fig.1 (top).
It is seen that $n(E)$ is drastically increasing for energies higher
than  $E_0\sim\Delta$. Additional calculations shows that
$n(E)$ do not depend on the number of sites $N$ for $E<E_0$
and is $\sim N$ for $E>E_0$. For all these reasons we conclude
 that states below $E_0$ are localised and states above $E_0$ are delocalised.

\section{Qualitative treatment}

On our opinion the main properties of the above model which are responsible
for the delocalisation are long range of intersite interaction $R$
and the fact that function $\rho(\varepsilon)$ differs from zero only
in the finite region.
For these reasons for the qualitative interpretation we apply the
following exactly solvable simple model of disordered system. Let
the radius of interaction  be infinite and write down the simplified
Hamiltonian in the form:

\begin{equation}
H_{r,r'}=\delta_{r,r'}\varepsilon_r+{v\over N},\hskip3mm r,r'=1,...,N
\end{equation}

Taking advantage of the coherent potential approximation
\cite{CPA,Pastur} one can show that the density of states for
Hamiltonians (1) and (5) is coincide in the limit
$R\rightarrow\infty, v_0 \rightarrow 0, 2Rv_0=v$. We
show below that the Hamiltonian (5) has one delocalised
and $N-1$ localised eigen functions. Consequently  at
least one delocalised function should appear in the
set of eigen functions of the Hamiltonian  (1)
 in the limit $R\rightarrow\infty$.
 The desire to see how this take place was the
 starting point for our study of the Hamiltonian (1) with
 $R>>1$. Now let us turn to the  proof of the above properties
of the Hamiltonian (5).

The equation for eigen vector ${\bf e}$ and eigen value $\lambda$
 of the Hamiltonian (5)  can be written in the form:

\begin{equation}
e_r={v\over N} \hskip2mm{S\over \lambda-\varepsilon_r}
\hskip10mm S\equiv\sum_{r=1}^N e_r
\end{equation}

(6) gives an explicit expression for the eigen vectors of (5) as
 a functions of $r$ and eigen number $\lambda$. By substituting $e_r$  in
 the formula for  $S$ one can obtain the equation for the eigen values $\lambda$:

\begin{equation}
 {1\over N} \sum_{r=1}^N {1\over \lambda-\varepsilon_r}
 \equiv \Gamma(\lambda)={1\over v}
 \end{equation}

For Anderson's model (3) all  quantities $\varepsilon_r$
 are differs from each other. For the graphical
treatment of (7) the qualitative form of $\Gamma(E)$
 function is presented on fig.2.
From fig.2 one can see that $N-1$ eigen values are belong to $[0,\Delta]$ and
 the last eigen value $E_m$ is not
belong to this interval and in the thermodynamic limit can be determined from the equation:
\begin{equation}
\Gamma(E) = \int{\rho(\varepsilon)d\varepsilon\over E-\varepsilon}
={1\over\Delta}\ln{E\over E-\Delta}= {1\over v}
\end{equation}

whence

\begin{equation}
E_m={\Delta\over 1-\exp -(\Delta/ v)}
\end{equation}

So in the thermodynamic limit $E_m$ is separated from any of $\varepsilon_r$
by finite interval. From (6) one can see that sharp extremums  of the
 wave function related to the localisation can
 appear if $\lambda\in[0,\Delta]$. $E_m$ do not
 belong to this interval and we come to the conclusion that the
 corresponding eigen vector
in the case of homogineous disorder is {\it delocalised}.
Now let us  show that all others eigen vectors are localised.
For this reason introduce the Green's function in t-representation
$\exp (it{\bf H})_{r,r}$ which describe the dynamics of the wave function
 on the site $r$ if it  was equal to 1 on this site at $t=0$.
If the finite part of eigen states of the Hamiltonian ${\bf H}$
is delocalised this function goes down to zero when
$t\rightarrow\infty$ and $N\rightarrow\infty$.
If Green's function do not decrease it means that the
main part of eigen vectors is localised and the part
of delocalised states is extremely small \cite{Pastur}.
It is convenient to introduce the Green's function in
E-representation:
\begin{equation}
\exp it{\bf H}={1\over 2\pi i}\int {\bf G}(E-i\delta) \exp (iEt)dE
\hskip10mm \delta\rightarrow +0, t>0
\end{equation}

In the case of Hamiltonian (5) the Dyson's series for ${\bf G}$:

\begin{equation}
G_{r,r'}=\delta_{r,r'}{1\over E-\varepsilon_r}+{1\over N}\hskip4pt
v {1\over E-\varepsilon_r}{1\over E-\varepsilon_r'}
+{1\over N}\hskip4pt v^2\Gamma(E){1\over E-\varepsilon_r}{1\over E-\varepsilon_r'}
+
\end{equation}
$$
+{1\over N} \hskip4pt v^3\Gamma(E)^2{1\over E-\varepsilon_r}{1\over E-\varepsilon_r'}
+...
$$
can be exactly summed and give the following expression for $G_{rr}$:

\begin{equation}
G_{r,r}=\bigg(E-\varepsilon_r-{1\over N}\hskip1mm{v\over 1-v g_r(E)}\bigg)^{-1}
\end{equation}

where

\begin{equation}
g_r(E)\equiv {1\over N}\sum_{l\neq r}^N {1\over E-\varepsilon_l}
\approx\Gamma(E)\approx{1\over\Delta}\ln{E\over E-\Delta}
\end{equation}

In the thermodynamic limit the term $\sim 1/N$ should be omitted and we
come to  the conclusion that the Green's function have a single pole $E=\varepsilon_r$.
This corresponds to the oscillations of the wave function with constant
amplitude and we can conclude  that the main part of states are localised.
It is easy to see that above described oscillations corresponds to
the wave function localised on the site $r$ and having an eigen value $\varepsilon_r+O(1/N)$.
From fig.2 one can see that there are $N-1$ eigen values of this
kind and we come to the conclusion that the Hamiltonian (5) have
$N-1$ localised states and one delocalised with eigen number $E_m$ (9).

Note that the appearance of the separated delocalised state for (5) is possible
 only if the distribution function $\rho(\varepsilon)$ is differ from zero in finite interval.
 For this reason  we expect that delocalisation in (1) is also possible if $\rho(\varepsilon)$
 is differ from zero in finite interval or at least goes down to zero rapidly enough.
 This statement confirms by calculations for Lloyd's model with
 $\rho(\varepsilon)=(1/\pi)\Delta/(\Delta^2+\varepsilon^2)$
 when no delocalisation was found.

The energy dependance of number of covered sites for Hamiltonian (5) is
 presented on fig.1 (bottom) for $v=20$. Other parameters are the
  same as for the top picture.
One can see that finiteness of the interaction radius  results in
 appearance of delocalised states in the gap $[\Delta, E_m]$
  but the boundary energy of spectrum and the mobility edge are
   the same for both Hamiltonians (1) and (5) and are equal
     $E_m$ (9) and $\Delta$ respectively.

\begin{figure}
\epsfxsize=400pt
\epsffile{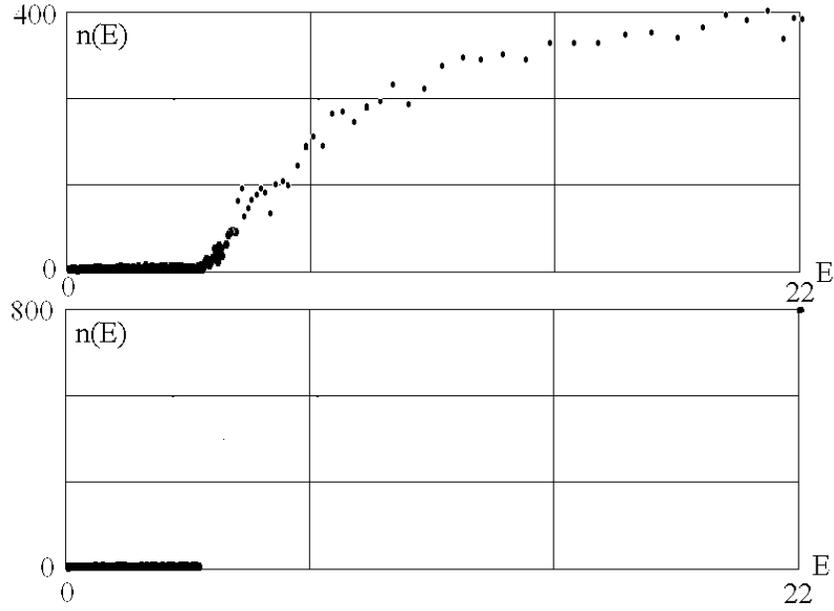}
\caption{Energy dependance of number of sites covered by the wave function
 for the Hamiltonian (1) with $N=1000$, $\Delta=4$,
   $w(r)=v_0\exp -|r/R|$,  $v_0=0.5$ , $R=20$  (top).
The same for the Hamiltonian (5) with $v=2Rv_0=20$ (bottom). }
\end{figure}

\begin{figure}
\epsfxsize=400pt
\epsffile{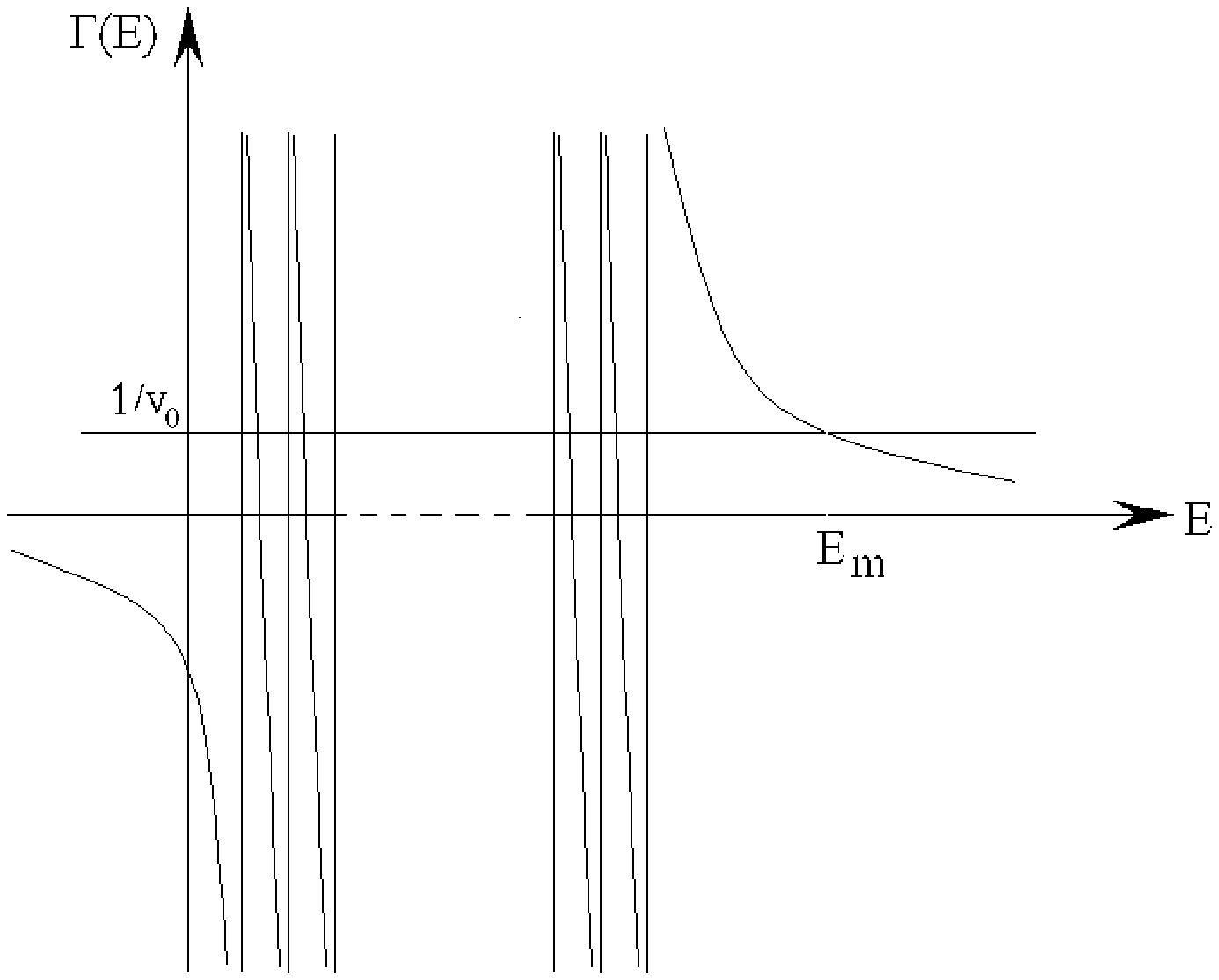}
\caption{}
\end{figure}


\begin{references}
\bibitem{Anderson} P.W.Anderson, Phys.Rev. {\bf 109}, 1492 (1958)
\bibitem{AM} arXiv: cond-mat/0303092 v2 9 Aug 2003
\bibitem{VM} Phys.Rev.Lett.,  A.Rodriguez, V.A.Malyshev, G.Sierra, M.A.
Anderson Transition in Low-Dimensional Disordered Systems Driven by Long-Range
Nonrandom Hopping, v90, n2 2003.
\bibitem{Pastur}
I.M.Lifshits, S.A.Gredeskul, and L.A.Pastur, Introduction in Theory of
Disordered Sysytems, Nauka, Moskow (1982)
\bibitem{CPA} Ved'aev A.V., Journal of Theoretical
and Mathematical Physics, 1977, v. 31, p. 392
\bibitem{Lloyd} J.Phys. C: Solid State Phys. 1969. V. 2. P. 1717.
\bibitem{Kozlov}arXiv: cond-mat/9909335
\end{references}
\end{document}